\documentclass{appolb}
\usepackage{graphicx}
\usepackage[utf8]{inputenc}
\usepackage[OT4]{fontenc}
\begin{document}
\title{Measurement of  proton-induced radiation in animal tissue%
\thanks{Presented at the XXXV Mazurian Lakes Conference on Physics, Piaski, Poland, September
3-9, 2017}%
}
\author{P. Sękowski, I. Skwira-Chalot, T. Matulewicz
\address{Faculty of Physics, University of Warsaw, PL-02093, Warsaw, Poland}
}

\maketitle
\begin{abstract}
Hadron therapy, because of the dosimetric and radiobiological advantages, is
more and more often used in tumour treatment.  This treatment method leads
also to the radioactive effects induced by energetic protons on nuclei. 
Nuclear reactions may lead to the production of radioactive isotopes. 
In the present experiment, two animal (human-like) tissue samples were irradiated
with 60 MeV protons.  Gamma-ray spectroscopy and lifetime measurements
allowed identifying isotopes produced during the irradiation, e.g. 
$^{18}$F and $^{34m}$Cl.  \\

\end{abstract}
  
\section{Introduction}
The present project is motivated by insufficient knowledge about long-lived
radioisotopes, which can be produced during proton therapy \cite{review,
janis}.  Except for the high linear transfer of energy, the efficiency of
particle therapy can also be augmented by induced radioactivity.  During
radioactive decays, different particles (which deposit energy in surrounding
tissues) are emitted and synergistic effect can occur.  The efficiency
of tumour cell killing by mixed radiation is higher than that for a separated
radiation.

The main goal of the project is an appraisal of dose from the induced
radioactivity deposited in irradiated and surrounding tissues.  Gamma-ray spectroscopy and lifetime measurements can be used to determine
the amount of isotopes produced during irradiation.  To provide the
consistency of the result, the Geant4/GATE \cite{gate, geant4} simulations
are used.

\section{Materials and Methods}

The experiment was performed at the Institute of Nuclear Physics of
the Polish Academy of Sciences in Cracow.  The proton beam accelerated to 60 MeV
(proton energy for eye therapy) was provided by the \mbox{AIC-144} isochronous cyclotron
and the samples were irradiated with doses in the range from 30 Gy
to 500 Gy.  To achieve a homogeneous distribution of the dose in the sample,
a technique called Spread Out of the Bragg Peak \cite{SOBP} was used.

Liver and bone samples were irradiated.  These samples are composed of
not only light nuclides like hydrogen, carbon or oxygen, but also of heavier
ones,
like potassium or iron.  Furthermore, those tissues have an ability to
accumulate much heavier elements.  The pig liver was chosen because of
its composition, which is similar to the human one, and its easy availability.  The
samples of bone were prepared from a beef meal with a few additional drops of
water.  Liver and bone samples were frozen using liquid nitrogen so that
they kept their shape during irradiation by a horizontal beam.

\section{Results}

 The energy spectrum of gamma rays emitted by the beef bone irradiated with
the dose equal to 250 Gy is presented in Fig.  1.  The spectrum was measured
using a HPGe detector.  There are several notable peaks: the 511 keV $\beta^+$
annihilation peak, 147 keV, 1157 keV and 2127 keV.  Except for the 1157 keV
line $\left(^{44}\textrm{Sc}\right)$, they originate from $^{34m}$Cl
\cite{baza}.

\begin{figure}[htb]
\centerline{%
\includegraphics[width=12.5cm]{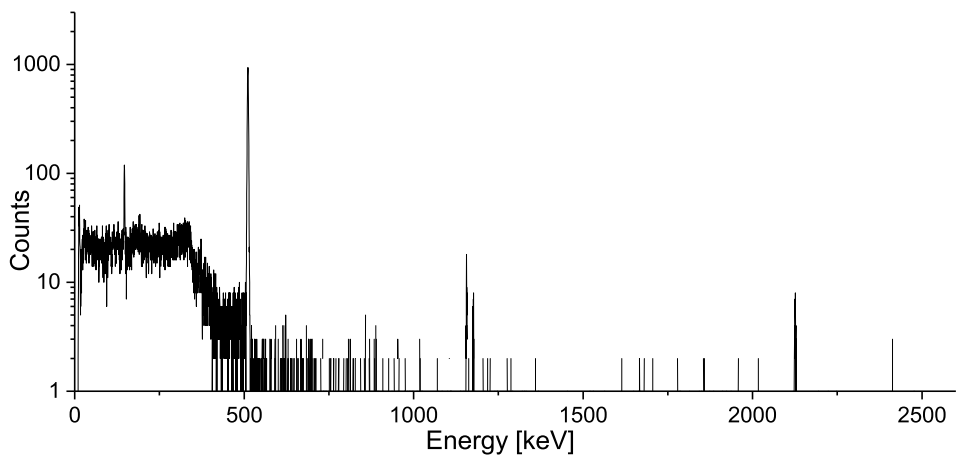}}
\caption{Gamma-ray energy spectrum of a bone irradiated with a dose of 250
Gy, measured with a HPGe detector placed in a low-background lead shield. The measurement started 2 hours after irradiation and lasted for 2 minutes.}
\label{Fig:Bone}
\end{figure}

Fig.  2 presents the gamma-ray energy spectrum of an irradiated pig liver
measured using a
LaBr$_3$ scintillator detector.  There are three notable peaks: the annihilation
peak, 683 keV and 1460 keV.  Three main sources of the annihilation peak are
$\beta^+$ decays of $^{11}$C, $^{13}$N and $^{18}$F, which is  confirmed by the time
spectrum exhibiting the three decay constants of those isotopes.  The line at the
energy of 1460 keV originates from the natural background ($^{40}$K).  The
peak with energy around 680 keV has no confirmed origin.  The most probable
source of this $\gamma$-ray  line is $^{204}$At because of the energy and
intensity of the gamma-ray line and a similar half-life.  For the further
calculation, it was assumed that this isotope is the source of the observed
radiation.

\begin{figure}[htb]
\centerline{%
\includegraphics[width=12.5cm]{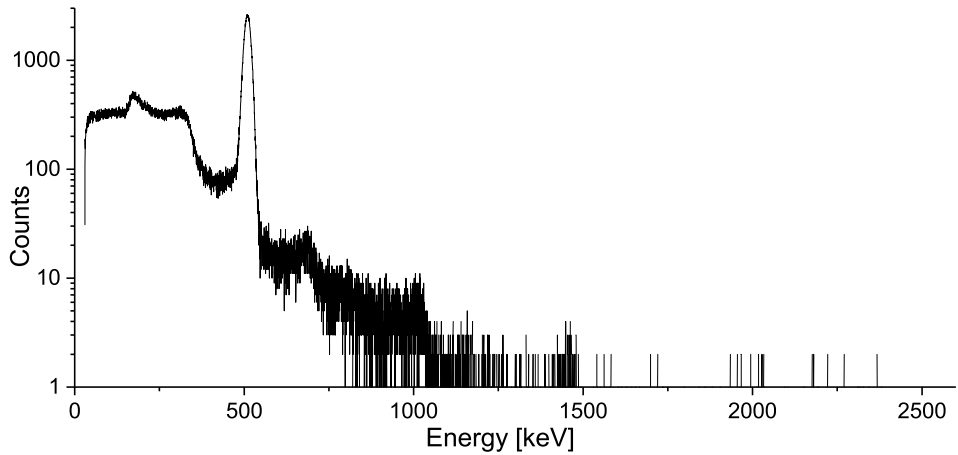}}
\caption{ Gamma-ray energy spectrum of liver irradiated with a dose of 500 Gy
measured with a LaBr$_3$ detector. The measurement started 10 minutes after irradiation and lasted for 100 seconds.}
\label{Fig:Liver}
\end{figure}

To estimate the dose from proton-induced radioactive isotopes, Monte
Carlo simulations were performed using the  Geant4/GATE package
\cite{gate,geant4}.  In the simulation, the radioactive isotopes were
located in the centre of a water sphere of 4 cm diameter.  Fig.  3 presents
examples of spatial dose distributions obtained for  two isotopes, $^{11}$C~(left) and
$^{34m}$Cl (right).

\begin{figure}[htb]
\centerline{%
\includegraphics[width=12.5cm]{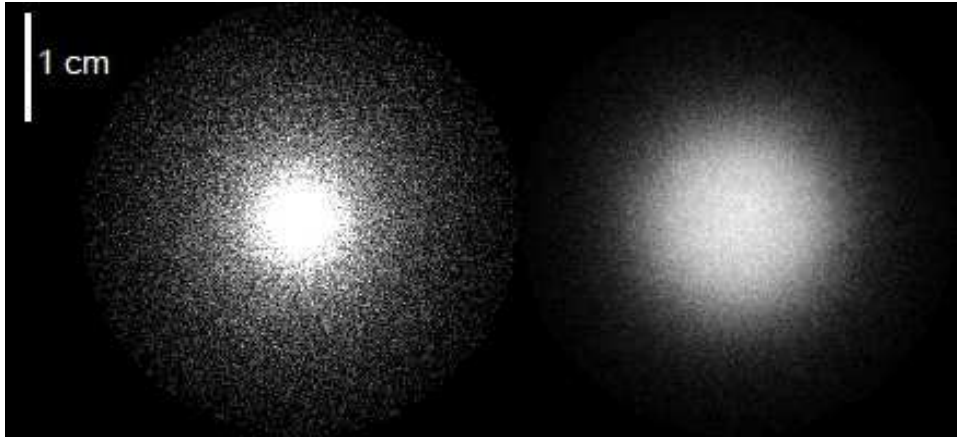}}
\caption{Spatial projections of dose distribution in water from point-like
sources of  $^{11}$C~(left) and $^{34m}$Cl (right).}
\label{Fig:Dist}
\end{figure} \newpage

In order to calculate the dose delivered to the surrounding tissue from the observed
radioisotopes, the irradiated sample volume and the total received dose were
taken into account (see Tab.  1).

\vspace{1cm}

\begin{table}[htb]
\centering
\renewcommand{\arraystretch}{1.2}
\caption{Dose from notable radioisotopes.}
\begin{tabular}{ccc}
\hline
Radioisotope & Dose [Gy$_{isotope}$/Gy$_{therapy}$ cm$^{3}$] & Tissue\\
\hline
$^{11}$C & 8.7 $\cdot$10$^{-9}$ & Liver\\ 
$^{13}$N & 1.9 $\cdot$10$^{-9}$& Liver\\ 
$^{18}$F & 4.3 $\cdot$10$^{-11}$& Liver\\ 
$^{34m}$Cl & 3.9 $\cdot$10$^{-9}$ & Bone\\ 
$^{204}$At & 7.8 $\cdot$10$^{-10}$ & Liver\\ 
\hline
\end{tabular}
\label{}
\end{table}

\section{Conclusions}
Based on the presented results, there is no indication that the induced
radioactivity,
created during eye proton therapy, changes significantly the global
therapeutic effects.  Some curious gamma-ray lines were observed, for
example that at 683 keV,
that should be studied further.  In order to test the local effects of the
induced radioactivity, radiobiological studies should be performed.

\section{Forthcoming Research}
During particle therapy of deeply located tumours, a more energetic proton
beam is used.  Therefore an extension of the present experimental
activities to higher beam energies is needed.  The experiment will be
continued also for other types of therapeutic beams, like carbon ions and
neutrons. 

\noindent
The last step of this project will explore the influence of radioactivity
induced in
the irradiated tissue on the surrounding non-irradiated cells.

\section*{\centering Acknowledgements }

The authors would like to thank Dr. Jan Swakoń and his team for performing irradiations and general support.
Several samples were measured in the laboratory of Prof. Jerzy Mietelski. We are grateful to him and his collaborators
for kind cooperation.


\begin{thebibliography}{1}

\bibitem{review}
M. Durante and H. Paganetti, 
{\em Nuclear physics in particle therapy: a review},
Rep. Prog. Phys. 79 (2016) 096702.

\bibitem{janis}
N.~Soppera, M.~Bossant, and E.~Dupont,
{\em JANIS 4: An Improved Version of the NEA Java-based Nuclear Data Information
System},
Nucl. Data Sheets, 120 (2014) 294.

\bibitem{gate}
S.~Jan {\em et~al.},
{\em GATE: a simulation toolkit for PET and SPECT}, 
Physics in Medicine and Biology, 49 (2004) 4543.

\bibitem{geant4}
S.~Agostinelli {\em et~al.},
{\em  Geant4 - a simulation toolkit}, Nucl. Instr. Meth. Phys. Res.  A 506
(2003) 250.

\bibitem{SOBP}
D. Yeung and J. Palta,
{\em Spread-Out Bragg Peak (SOBP)}, pages 809--810,
Springer Berlin Heidelberg, Berlin, Heidelberg, 2013.

\bibitem{baza}
{Nuclear Data Center at Korea Atomic Energy Research Institute}, 
{\tt http://atom.kaeri.re.kr/} (accessed 1 Aug 2017)

\end{thebibliography}
\end{document}